\newcommand{\DoPrePrint}{0} 

\ifnum\DoPrePrint=1
  \RequirePackage{lineno} 

  \documentclass[preprint, review, 1p, number, sort&compress]{elsarticle}
  \linenumbers
  \usepackage{lineno}
\else
  \documentclass[preprint, twocolumn, 5p, number, sort&compress]{elsarticle}
\fi

\usepackage{amsmath}   
\usepackage{graphicx}  
\usepackage{xspace}
\usepackage{units}
\usepackage{gensymb}
\usepackage{hyperref}
\usepackage{tabularx}

\newcommand{\minerva}{MINERvA\xspace}
\newcommand{\minos}{MINOS\xspace}

\newcommand{\footnoteremember}[2]{\footnote{#2}\newcounter{#1}\setcounter{#1}{\value{footnote}}}

\newcommand{\PLsupp}{0}   
\ifnum\PLsupp=0
  \newcommand{\SuppLocation}{in the Appendix}
\else
  \newcommand{\SuppLocation}{} 
\fi

\newif\ifpdf
\ifx\pdfoutput\undefined
   \pdffalse
\else
   \pdfoutput=1
   \pdftrue
\fi
\ifpdf
   \usepackage{graphicx}
   \usepackage{epstopdf}
\else
   \usepackage{graphicx}
\fi

\begin{document}
\begin{frontmatter}

\ifnum\DoPrePrint=1
\linenumbers
\fi
\title{Single neutral pion production by charged-current $\bar{\nu}_\mu$ interactions on hydrocarbon at $\langle E_\nu \rangle = $ 3.6 GeV}

\newcommand{\Rutgers}{Rutgers, The State University of New Jersey, Piscataway, New Jersey 08854, USA}
\newcommand{\Hampton}{Hampton University, Dept. of Physics, Hampton, Virginia 23668, USA}
\newcommand{\Dortmund}{Institute of Physics, Dortmund University, 44221, Germany }
\newcommand{\Otterbein}{Department of Physics, Otterbein University, 1 South Grove Street, Westerville, Ohio, 43081 USA}
\newcommand{\JMU}{James Madison University, Harrisonburg, Virginia 22807, USA}
\newcommand{\Florida}{University of Florida, Department of Physics, Gainesville, FL 32611, USA}
\newcommand{\UCIrvine}{Department of Physics and Astronomy, University of California, Irvine, Irvine, California 92697-4575, USA}
\newcommand{\CBPF}{Centro Brasileiro de Pesquisas F\'{i}sicas, Rua Dr. Xavier Sigaud 150, Urca, Rio de Janeiro, RJ, 22290-180, Brazil}
\newcommand{\PUCP}{Secci\'{o}n F\'{i}sica, Departamento de Ciencias, Pontificia Universidad Cat\'{o}lica del Per\'{u}, Apartado 1761, Lima, Per\'{u}}
\newcommand{\INRM}{Institute for Nuclear Research of the Russian Academy of Sciences, 117312 Moscow, Russia}
\newcommand{\Jlab}{Jefferson Lab, 12000 Jefferson Avenue, Newport News, Virginia 23606, USA}
\newcommand{\Pittsburgh}{Department of Physics and Astronomy, University of Pittsburgh, Pittsburgh, Pennsylvania 15260, USA}
\newcommand{\Guanajuato}{Campus Le\'{o}n y Campus Guanajuato, Universidad de Guanajuato, Lascurain de Retana No. 5, Col. Centro. Guanajuato 36000, Guanajuato M\'{e}xico.}
\newcommand{\Athens}{Department of Physics, University of Athens, GR-15771 Athens, Greece}
\newcommand{\Tufts}{Physics Department, Tufts University, Medford, Massachusetts 02155, USA}
\newcommand{\WM}{Department of Physics, College of William \& Mary, Williamsburg, Virginia 23187, USA}
\newcommand{\FNAL}{Fermi National Accelerator Laboratory, Batavia, Illinois 60510, USA}
\newcommand{\Purdue}{Department of Chemistry and Physics, Purdue University Calumet, Hammond, Indiana 46323, USA}
\newcommand{\MCLA}{Massachusetts College of Liberal Arts, 375 Church Street, North Adams, Massachussetts 01247, USA}
\newcommand{\UMD}{Department of Physics, University of Minnesota -- Duluth, Duluth, Minnesota 55812, USA}
\newcommand{\Northwestern}{Northwestern University, Evanston, Illinois 60208, USA}
\newcommand{\UNI}{Universidad Nacional de Ingenier\'{i}a, Apartado 31139, Lima, Per\'{u}}
\newcommand{\Rochester}{University of Rochester, Rochester, New York 14610 USA}
\newcommand{\Austin}{Department of Physics, University of Texas, 1 University Station, Austin, Texas 78712, USA}
\newcommand{\USM}{Departamento de F\'{i}sica, Universidad T\'{e}cnica Federico Santa Mar\'{i}a, Avenida Espa\~{n}a 1680 Casilla 110-V, Valpara\'{i}so, Chile}
\newcommand{\Geneva}{University of Geneva, Geneva, Switzerland}
\newcommand{\Chicago}{Enrico Fermi Institute, University of Chicago, Chicago, Illinois 60637 USA}
\newcommand{\bmeThanks}{Now at SLAC National Accelerator Laboratory, Stanford, California 94309 USA}
\newcommand{\LazaThanks}{Also at Department of Physics, University of Antananarivo, Madagascar}
\newcommand{\ticeThanks}{Now at Argonne National Laboratory, Argonne, Illinois 60439, USA }
\newcommand{\janThanks}{Also at Institute of Theoretical Physics, Wroc\l aw University, Wroc\l aw, Poland}

\author[Rutgers]{T.~Le\corref{corrauthor}} 
\ifnum\PLsupp=1
  \ead{ltrung@physics.rutgers.edu}  
\fi
\author[CBPF]{J.L.~Palomino}                    
\author[WM,PUCP]{L.~Aliaga}
\author[Tufts]{O.~Altinok}                       
\author[Rochester]{A.~Bercellie}  
\author[Rochester]{A.~Bodek} 
\author[Geneva]{A.~Bravar}                        
\author[USM]{W.K. Brooks}                        
\author[INRM]{A.~Butkevich}                     
\author[CBPF,FNAL]{D.A.~Martinez~Caicedo}                   
\author[CBPF]{M.F.Carneiro}
\author[Hampton]{M.E.~Christy}                     
\author[Rochester]{J.~Chvojka}                       
\author[CBPF]{H.~da~Motta}                      
\author[WM]{J.~Devan}                         
\author[Pittsburgh]{S.A.~Dytman}                      
\author[PUCP]{G.A.~D\'{i}az~}                   
\author[Pittsburgh]{B. Eberly\fnref{bmeThanks}}
\author[Guanajuato]{J.~Felix}                         
\author[Northwestern]{L.~Fields}                        
\author[Rochester]{R.~Fine}                          
\author[PUCP]{A.M.~Gago}                        
\author[Tufts]{H.~Gallagher}                     
\author[UMD]{R.~Gran}                          
\author[FNAL]{D.A.~Harris}                      
\author[Rochester,Guanajuato]{A.~Higuera}
\author[CBPF,UNI]{K.~Hurtado}
\author[WM]{M.~Kordosky}                      
\author[MCLA]{E.~Maher}                         
\author[Rochester]{S.~Manly}                         
\author[Tufts]{W.A.~Mann}                        
\author[Rochester]{C.M.~Marshall} 
\author[Rochester,FNAL]{K.S.~McFarland}
\author[Pittsburgh]{C.L.~McGivern}                    
\author[Rochester]{A.M.~McGowan}                     
\author[USM]{J.~Miller}                        
\author[FNAL]{J.G.~Morf\'{i}n}                  
\author[Florida]{J.~Mousseau}                      
\author[WM]{J.K.~Nelson}                      
\author[WM]{A. Norrick}
\author[FNAL]{J.~Osta}                          
\author[Pittsburgh]{V.~Paolone}                       
\author[Rochester]{J.~Park}                          
\author[Northwestern]{C.E.~Patrick}                     
\author[FNAL,Rochester]{G.N.~Perdue}
\author[FNAL]{L.~Rakotondravohitra\fnref{LazaThanks}} 
\author[Rutgers]{R.D.~Ransome}                     
\author[Florida]{H.~Ray}                           
\author[Pittsburgh]{L.~Ren}                           
\author[Rochester]{P.A.~Rodrigues}                   
\author[Rochester]{D. Ruterbories}
\author[Northwestern]{H.~Schellman}                     
\author[Chicago,FNAL]{D.W.~Schmitz}
\author[FNAL]{J.T.~Sobczyk\fnref{janThanks}} 
\author[UNI]{C.J.~Solano~Salinas}              
\author[Otterbein]{N.~Tagg}                          
\author[Rutgers]{B.G.~Tice\fnref{ticeThanks}}   
\author[Guanajuato]{E. Valencia}
\author[Hampton]{T.~Walton}                        
\author[Rochester]{J.~Wolcott}                       
\author[CBPF]{H. Yepes-Ramirez}
\author[Guanajuato]{G.~Zavala}                        
\author[WM]{D.~Zhang}                         
\author[UCIrvine]{B.P.Ziemer}

\cortext[corrauthor]{Corresponding author\ifnum\PLsupp=1. Phone: +1 631-371-9595\fi}

\address[Rutgers]{\Rutgers}
\address[CBPF]{\CBPF}
\address[WM]{\WM}
\address[Tufts]{\Tufts}
\address[Rochester]{\Rochester}
\address[Geneva]{\Geneva}
\address[USM]{\USM}
\address[INRM]{\INRM}
\address[Hampton]{\Hampton}
\address[Pittsburgh]{\Pittsburgh}
\address[PUCP]{\PUCP}
\address[Guanajuato]{\Guanajuato}
\address[Northwestern]{\Northwestern}
\address[UMD]{\UMD}
\address[FNAL]{\FNAL}
\address[MCLA]{\MCLA}
\address[Florida]{\Florida}
\address[Chicago]{\Chicago}
\address[UNI]{\UNI}
\address[Otterbein]{\Otterbein}
\address[UCIrvine]{\UCIrvine}

\fntext[bmeThanks]{\bmeThanks} 
\fntext[LazaThanks]{\LazaThanks} 
\fntext[janThanks]{\janThanks}  
\fntext[ticeThanks]{\ticeThanks} 


\begin{abstract}

Single neutral pion production via muon antineutrino charged-current interactions in
plastic scintillator (CH) is studied using the \minerva detector 
exposed to the NuMI low-energy, wideband antineutrino beam at Fermilab.  
Measurement of this process constrains models of neutral pion 
production in nuclei, which is important because the
neutral-current analog is a background
for $\bar{\nu}_e$ appearance oscillation experiments.  The
differential cross sections for $\pi^0$ momentum and production angle, 
for events with a single observed $\pi^0$ and no charged pions, are presented
and compared to model predictions. These results comprise the first measurement of the
$\pi^0$ kinematics for this process.

\end{abstract}


\end{frontmatter}

\section{Introduction}

Neutrino- and antineutrino-induced interactions at energies of
a few GeV are a proving ground for weak interaction
phenomenology in nuclei~\cite{Alvarez-Ruso:2014bla}.
Measurements in this energy range are important because 
neutrino oscillation experiments~\cite{Ayres:2004js,LBNE} need 
detailed understanding of the large variety of processes allowed.  
As a result, there is a growing body of new high-quality measurements for 
neutrino-nucleus
interactions~\cite{AguilarArevalo:2007ab,Lyubushkin:2008pe,nubarprl,miniboone_piprod}. 
In particular, neutrino charged-current neutral pion 
production has become an important benchmark providing new challenges 
to theories describing this process~\cite{gibuu-pi,valencia-pi,Rodrigues:2014jfa,Yu:2014yja}.

Most of the published data for pion production in nuclei uses 
neutrino beams.  Neutral pion production in nuclei for anti-neutrinos,
however, is much less studied.
Only one data point for this channel in the few-GeV energy range exists 
in the literature: 
a measurement of $\bar{\nu}_\mu p \rightarrow \mu^+ n \pi^0$
from SKAT in a heavy liquid (CF$_3$Br) bubble chamber based on 20 events
at an average neutrino energy of 7 GeV~\cite{Grabosch89}. 
Measurements of $\pi^0$ production by neutrinos have been made in 
deuterium bubble chambers~\cite{Allasia90, Barish79, Radecky82, Kitagaki86}
 and more recently on nuclear targets in the 0.1 - 1 GeV energy range
using 
the MiniBooNE detector~\cite{AguilarArevalo:2008qa} with a mineral oil (CH$_2$) target, and
using the SciBar detector in K2K and SciBooNE 
experiments~\cite{Mariani:2007,Kurimoto:2009} with plastic scintillator (CH).
These recent measurements, as well as data on charged pion 
production~\cite{miniboone_piprod,Eberly:2014mra} and 
neutral pion production~\cite{Augil11} have been difficult 
for event generators and theoretical calculations to 
describe accurately~\cite{gibuu-pi,valencia-pi,Rodrigues:2014jfa,Yu:2014yja}.
Every prediction must have a model for $\pi^0$ production from
nucleons; all use isospin decompositions within
a helicity formalism that are tuned to available data~\cite{Rein:1980wg}.

Charged-current single $\pi^0$ ($1\pi^0$) production in the few-GeV region
is modeled both as decays of nucleon resonances (most strongly
the $\Delta(1232)$) and nonresonant processes. The production 
in nuclei can be
either direct, through the reaction $\bar{\nu}_\mu p \rightarrow \mu^{+} n
\pi^0$, or indirect, for example, through charge exchange (CEX) of a charged pion in the nucleus, 
$p\pi^-\rightarrow n\pi^0$ or $n\pi^+\rightarrow p\pi^0$.
New data will provide useful tests  
of both neutrino-induced resonance production and final state 
interaction (FSI) models. 

Neutrino interaction measurements are also important to the analysis of neutrino
oscillation experiments~\cite{Abe:2011ks,Ayres:2004js,AguilarArevalo:2010wv,LBNE}. 
These experiments require the neutrino flavor be
identified and the neutrino energy to be reconstructed on an event-by-event basis.
An accurate modeling of these particles requires knowledge of both the underlying
neutrino-nucleon interactions and of the final-state modifications that arise within the
target nuclei (such as carbon) of which the massive oscillation detectors are comprised.
Pion production is a source of backgrounds and systematic uncertainties in neutrino oscillation experiments. 
Neutral-current $\pi^0$ production, for example, is a dominant background in
$\nu_e$ ($\bar{\nu}_e$) appearance experiments because the $\pi^0$ can mimic 
a final state electron (positron).
In addition, experiments that reconstruct neutrino energy by 
identifying quasielastic events,
~$\nu_\ell N(n) \rightarrow \ell^{-} p$~
or~$\bar{\nu}_\ell N(p) \rightarrow \ell^{+} n$, 
interactions in which a pion is produced but then absorbed in the target 
nucleus can be mistaken for quasielastic signal and yield an incorrect estimate of the 
incident neutrino energy.

New measurements of $1\pi^0$ production by 
charged-current $\bar{\nu}_\mu$ interactions in plastic scintillator (CH) using
the \minerva detector are presented.
Flux-integrated single differential cross sections as a function of $\pi^0$ momentum and production angle for events with a
single observed $\pi^0$ and no $\pi^\pm$ exiting the interaction nucleus have
been measured and are compared to
predictions from the GENIE~\cite{Andreopoulos201087}, NuWro~\cite{Juszczak:2005zs,Golan:2012rfa}, 
and NEUT~\cite{Hayato:2009zz} event generators.

\section{Experiment}

The data presented here were taken using the \minerva detector and the wideband antineutrino beam produced by the NuMI beamline
in the low-energy mode~\cite{Anderson:1998zza} with a mean energy of $3.6$~GeV.
The antineutrino flux is estimated from a simulation of the neutrino beamline based on Geant4~\cite{Agostinelli2003250,1610988}, with
hadron production in the simulation constrained by proton-carbon external data~\cite{Alt:2006fr,Barton:1982dg,Lebedev:2007zz}.
The \minerva detector consists of a fully active region of scintillator strips surrounded on the sides 
and downstream end by electromagnetic and hadronic calorimeters, and is described in detail in Ref.~\cite{minerva_nim}.
There are three orientations (views) for the strips (X,U,V), offset by $60^\circ$ from each other, which enable 
three-dimensional reconstruction of particle trajectories.
The X view is sampled twice as often as the other views.
The downstream edge of the \minerva detector is located \unit[2]{m} upstream of the MINOS Near Detector, 
a magnetized iron spectrometer~\cite{Michael:2008bc} used in this analysis 
to reconstruct the momentum and charge of muons.
The transport of particles from neutrino interactions in the detector is simulated by a Geant4-based program.
The readout simulation is tuned so that both the photostatistics and the reconstructed energy deposited
by momentum-analyzed through-going muons agree between data and the simulation. The detector 
simulation of single particle responses is validated using testbeam data taken with a scaled-down 
version of the MINERvA detector~\cite{Aliaga:2015aqe}.

Neutrino interactions are simulated using the GENIE 2.6.2 neutrino event generator. 
Details concerning GENIE and its associated parameters are described in Ref.~\cite{Andreopoulos201087}.
For baryon resonance production, the formalism of Rein-Sehgal~\cite{Rein:1980wg} is used with modern resonance properties~\cite{pdg}.
Non-resonant pion production is simulated using the Bodek-Yang model~\cite {Bodek:2004pc} and is constrained below $W=1.7$~GeV 
by neutrino-deuterium bubble chamber data~\cite{Radecky:1981fn,Kitagaki:1990vs}.
Pion and nucleon FSI are modeled in GENIE using a parameterized intranuclear cascade model, 
with the full cascade being represented by a single interaction.  For all models fitting data for light nuclei such as carbon, 
the pion most often has one interaction as it propagates through the nucleus.  
For each interaction, choice of the channel (e.g. charge exchange) is based on
total cross section data and calculations~\cite{Lee:2002eq,Ashery:1981tq} and the kinematics and multiplicity of the final state are taken from fits to more detailed data.
At the energies important for this measurement, the hadron-nucleus cross sections have large uncertainties.
For example, the cross section of $\pi^-\rightarrow \pi^0$ CEX on carbon has an uncertainty of about 50\%.
Reaction cross sections for $\pi^0$ are estimated from measured cross sections for $\pi^\pm$ scattering using isospin symmetry or theoretical models.
The most significant advantages of the single interaction used in GENIE's FSI model
are the ability to exactly reweight and to characterize each event with a single final state channel.
The uncertainties in the FSI model are evaluated by varying its strength within previously 
measured hadron-nucleus cross-section uncertainties.

This analysis uses data taken between October 2009 and February 2012 with $2.01\times 10^{20}$ protons on target (POT) in the $\bar{\nu}_\mu$ mode. 
About half of the exposure ($0.945\times 10^{20}$ POT) was taken during construction with the downstream half of the detector.
In this period the ArgoNeuT detector~\cite{Argoneut} was situated between the \minerva and MINOS Near detectors. 
Because the two sub-samples of the data have different efficiencies, they were analyzed separately and their results combined.

\section{Event reconstruction and selection} 

The \minerva\ detector records the charge and time of energy
depositions (hits) in each scintillator strip.  Hits are first grouped
in time and then clusters of energy are formed by spatially grouping
adjacent hits in each scintillator plane.  Clusters with energy
more than $\unit[1]{MeV}$ are then matched among the three views to create a
track. The per-plane position resolution is 2.7~mm and the
angular resolution of the muon track is better than
10~mrad~\cite{minerva_nim} in each view. The $\mu^+$ is identified by matching a
track that exits
the back of \minerva with a positively-charged track entering the
front of \minos. The reconstruction of the muon in the MINOS spectrometer gives
a typical momentum resolution of $11\%$. 
Event pile-up causes a decrease in the muon track reconstruction efficiency. 
This effect is studied in both \minerva and \minos by projecting tracks found in
one of the detectors to the other and measuring the mis-reconstruction rate. 
This results in a $-4.4\%$ ($-1.1\%$) correction to the simulated efficiency for
muons below (above) 3 GeV/c.

The event vertex, defined as the most upstream cluster on the muon track, is restricted to be
within the central 108 planes of the scintillator tracking region and
no closer than \unit[22]{cm} to any edge of the planes. These
requirements define a fiducial volume with a mass of \unit[5.47]{metric tons}.
Due to the requirement that the $\mu^+$ is tracked in MINOS for charge and momentum
measurement, the detection efficiency has a strong dependence on muon angle $\theta_\mu$ and
momentum $|\vec{p}_\mu|$, which drops to zero for  events with  $\theta_\mu$ greater than
$\unit[20]{degrees}$ or $|\vec{p}_\mu|$ less than $\unit[1.0]{GeV/c}$.
The corrections for the angle and momentum efficiency are estimated
from the event simulation. Only events with a single track at the vertex,
 the $\mu^+$ matched to MINOS, are used in order to reject events including
 charged pion production. 
Accepted events are passed to the $\pi^0$ reconstruction.  

The neutral pion has a lifetime of $\unit[8.52\times10^{-17}]{s}$ and decays
into two photons with a branching ratio of 98.8\%~\cite{pdg}, so the two photons appear
to come from the event vertex. 
Plastic scintillator has a radiation length $X_0\sim\unit[40]{cm}$, which
allows the two photons to convert by $e^+e^-$ pair production or Compton
scattering 
far away from the vertex, thus producing isolated energy deposits. 
The photons can also have significant energy leakage or escape the detector
without conversion. 
Furthermore, energy deposits produced by neutrons from the same neutrino
interaction can be mistaken for low-energy photon conversions, further
complicating the pattern of energy deposits. 
The $\pi^0$ reconstruction must correctly group these energy deposits into
the two photons.

The reconstruction can be separated into two steps: pattern recognition and
kinematic reconstruction. The pattern recognition builds upon the knowledge of the vertex
location of the event. In the X view, clusters that are close in polar angle with respect to
the vertex but can be
separated in radial distance from the vertex are grouped into photon candidates. 
Then, for each candidate, clusters in the U and V views consistent with the stereo condition are added. 
Photon candidates must have clusters from at least two views
for three-dimensional direction reconstruction.

In the second step, photon position, direction, and energy are determined from
the clusters that have been assigned to the photons. 
The photon direction is reconstructed from the cluster energy-weighted slopes in each view.
The photon vertex is defined by the closest cluster to the event
vertex on the photon direction axis. 
The photon energies are reconstructed by calorimetry using calibration constants 
determined from the simulation. 
The overall calibration constant that sets the absolute energy scale is
determined by matching the peak in the $\gamma\gamma$ invariant mass
distribution to the $\pi^0$ nominal mass of
$\unit[134.97]{MeV/c^2}$~\cite{pdg}. 
This procedure is done separately for data and simulation which enables
correction for a difference in energy scales of 5\% between the data and simulation. 
Finally, the $\pi^0$ momentum is calculated from momentum conservation, $\vec p_{\pi^0}
= \vec{k}_1 + \vec{k}_2$, where $\vec{k}_i$ 
are reconstructed photon momenta. 
The $\pi^0$ is reconstructed with a 25\% energy resolution and
$\unit[3.5]{degrees}$ angular resolution in each view. The incoming neutrino energy is
reconstructed from the $\mu^+$ and $\pi^0$ 4-momenta using
\begin{eqnarray}
 E_\nu^{rec} &=& E_\mu + E_{\pi^0} + T_n \\ \nonumber
 T_n &=& \frac{1}{2}\frac{[(E_\mu-p_\mu^\parallel)+(E_{\pi^0}-p_{\pi^0}^\parallel)]^2 + (\vec{p}_\mu^\perp + \vec{p}^\perp_{\pi^0})^2}
 {m_N-(E_\mu-p_\mu^\parallel)-(E_{\pi^0}-p_{\pi^0}^\parallel)},
\end{eqnarray}
where $\vec{p}^\perp, p^\parallel$ are the transverse and longitudinal components of momentum, respectively. 
It is assumed that the initial nucleon is at rest and that the $\pi^0$ is produced
together with a nucleon. The neutrino energy is reconstructed with 10\% resolution.
The $\gamma\gamma$ invariant mass $m_{\gamma\gamma}$ is reconstructed from the photon energies $E_{1,2}$ and
the separation angle $\theta_{\gamma\gamma}$ between the two photons using
\begin{equation}
m_{\gamma\gamma}^2=2E_1E_2(1-\cos\theta_{\gamma\gamma}). 
\end{equation}

The two reconstructed photons are required to convert at least $\unit[15]{cm}$ ($0.36X_0$) away
from the vertex to further reject charged-pion backgrounds. 
In addition, it is required that $E_\nu^{rec}$ is between 1.5 and 20 GeV.
The lower energy cut is needed due to MINOS acceptance while the upper cut is to reduce flux uncertainties.
Finally, it is required that $m_{\gamma\gamma}$ is between
$\unit[75]{MeV/c^2}$ and $\unit[195]{MeV/c^2}$. 
The selected sample has 1304 events. 
The total selection efficiency is 6\% and
purity 55\%, according to the simulation. 
The background is dominated (70\% of the total background) by events with at
least one $\pi^0$ in the detector. 
Half of this is due to multi-pion production, $\pi^0+\pi^{\pm}$,
where the $\pi^{\pm}$ is not tracked, while 
the other half has a secondary $\pi^0$ produced by $\pi^-\rightarrow\pi^0$ CEX 
or nucleon scattering in the detector but outside the primary
interaction nucleus. 
The non-$\pi^0$ background is mostly due to energy deposits by $\pi^-$ and neutrons
which are mistakenly
identified as photons, and accounts for the remaining 30\% of the total
background.

\begin{figure}[!htpb]
\centering
\ifnum\PLsupp=0
  \includegraphics[width=1.0\columnwidth]{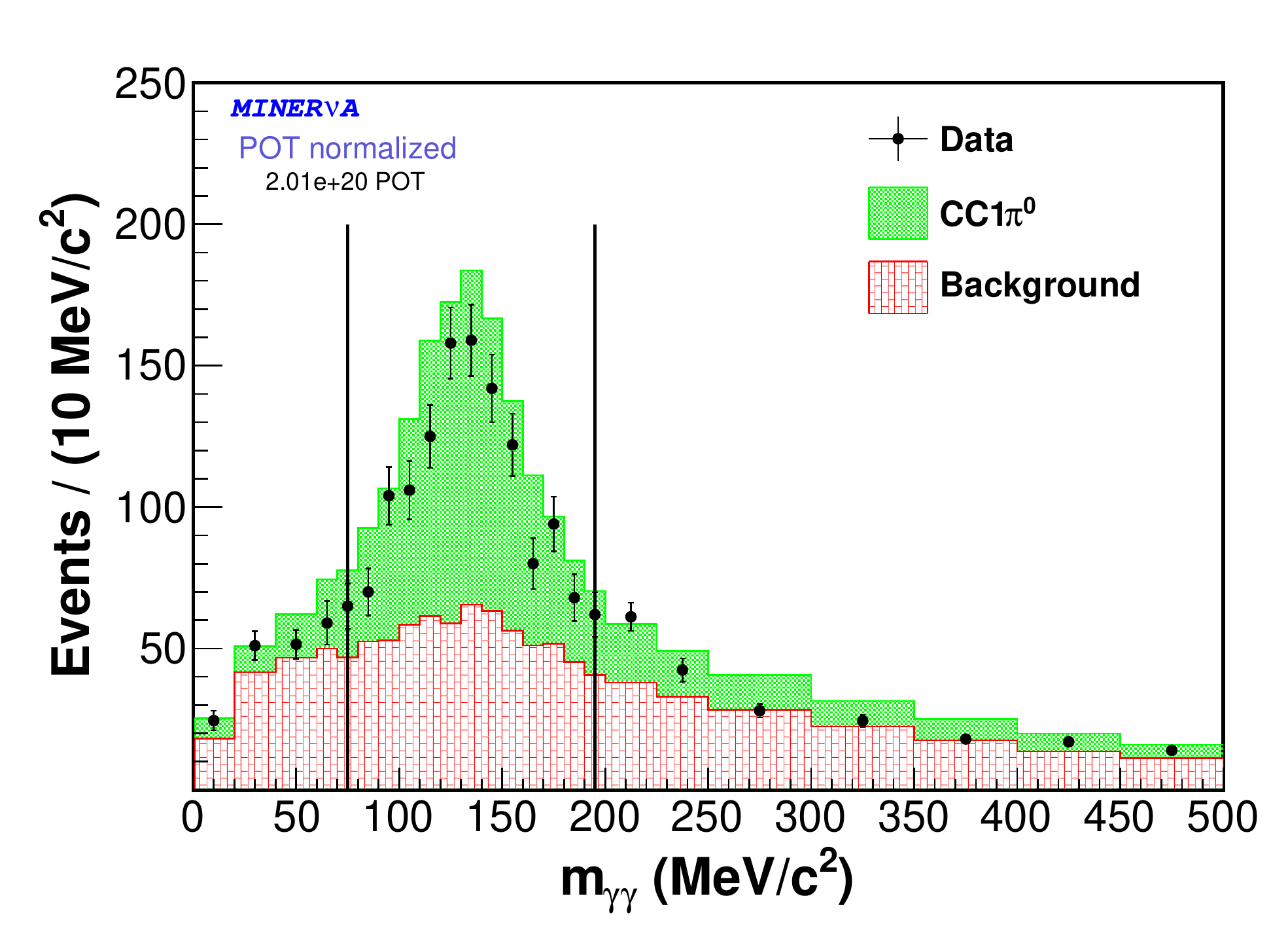}
\else
  \includegraphics[width=1.0\columnwidth]{mgg-data-mc-breakdown-pot.pdf}
\fi

\vspace{-7pt}

\caption{Distribution of the invariant mass of the $\gamma\gamma$ pair.
Data are shown as solid circles with statistical error bars. 
The shaded histograms show the Monte Carlo predictions for CC1$\pi^0$ signal (top) and for background (bottom). 
The signal is defined as antineutrino charged-current events with single $\pi^0$ and no $\pi^\pm$ escaping 
the nucleus. The background is anything else that is not signal. 
The vertical lines indicate the invariant mass cut, $\unit[75]{MeV/c^2} < m_{\gamma\gamma} < \unit[195]{MeV/c^2}$.}
\label{fig:invariant-mass}
\end{figure}

Figure \ref{fig:invariant-mass} shows the $m_{\gamma\gamma}$
distributions for both data and simulation of the selected sample before the
invariant mass cut. 
There is a clear peak centered around the $\pi^0$ nominal mass in both data and
simulation. 
The distribution from simulation is broken down into signal and background
components. The signal is defined as antineutrino 
charged-current events with single $\pi^0$ and no $\pi^\pm$ escaping 
the nucleus. The background is anything else that is not signal. By this definition, 
it is possible for signal events to have the $\pi^0$s mis-reconstructed from non-$\pi^0$ energy 
deposits. The signal events at high $m_{\gamma\gamma}$, outside the signal mass width, have one or 
both candidate photons reconstructed from neutron energy deposits. 
The same mis-reconstruction happens to signal events in the selected sample but at a smaller rate (14\%). 
The $\pi^0$ momentum and angular shapes of these events are found to be similar to the rest of the signal events.
The enhancement in the background distribution around the $\pi^0$
mass is due to background events with at least one $\pi^0$ in the
detector. 

After event selection, the selected
sample still has substantial background to be subtracted statistically
from the total distribution for each observable. 
The background distribution for each observable is estimated from the simulation
with the total background rate constrained by data. 
This significantly reduces the uncertainties in the estimated background. 
The background rate is determined from a binned extended maximum likelihood
fit of an invariant mass model to the data~\cite{Barlow:1993dm}. 
The invariant mass model $M(m_{\gamma\gamma})$ is a two-component model
constructed from the $m_{\gamma\gamma}$ distributions of signal and background
events,
\begin{equation}
M(m_{\gamma\gamma})=N_{sig}M_{sig}(m_{\gamma\gamma}) +
N_{bkg}M_{bkg}(m_{\gamma\gamma}),
\end{equation}
where $M_{sig}(m_{\gamma\gamma}), M_{bkg}(m_{\gamma\gamma})$ are the shapes of
the signal and background $m_{\gamma\gamma}$ distributions from the simulation, respectively. 
The expected numbers of signal and background events $N_{sig},N_{bkg}$ in the
range 0-500 MeV/$c^2$ are the parameters determined from the fit. 
After the fit, a background rate of 541$\pm$37 events is obtained by integrating
the curve $N_{bkg}M_{bkg}(m_{\gamma\gamma})$ over the mass peak region from
$\unit[75]{MeV/c^2}$ to $\unit[195]{MeV/c^2}$, the same range as required by the event selection.
The fit reduces the background normalization by a factor of 0.8 compared to the simulation prediction.

The estimated background is subtracted from the total distribution. This
background-subtracted data distribution is unfolded to account for detector
resolution effects 
using a Bayesian unfolding method~\cite{agostini:1995} with two iterations. 
The migration matrix used in the unfolding is obtained from the simulation. 
The reconstruction efficiency and acceptance corrections are made by dividing by
an efficiency curve derived from the simulation. 
Dividing this corrected distribution by the integrated flux
($2.5\times10^{-8}\bar{\nu}_\mu/\text{cm}^2/\text{POT}$) of antineutrinos with
energies in the range 1.5-20 GeV, 
and the number of nucleons in the fiducial volume ($3.30\times10^{30}$) gives
the differential cross section.


The total uncertainties on the measured cross sections are 20-25\% with comparable 
contributions from statistical and systematic uncertainties. The largest systematic uncertainty 
is due to the integrated flux uncertainty at 10\%. A detailed discussion of the flux systematic 
uncertainty is presented in Ref.~\cite{nubarprl}. The next largest contribution to the 
systematic uncertainty is the 8\% normalization uncertainty from the background fit. 
The other smaller contributions include the neutrino cross-section models, FSI models, 
and detector simulation systematics. These systematic uncertainties enter the measured 
cross sections through background subtraction, detector resolution, and efficiency corrections. 
The neutrino cross-section model and FSI model uncertainties are evaluated by GENIE. 
The systematic uncertainties on the estimated background is small since the background rate is constrained by data. 
Systematic errors in the detector response must be independently estimated.
The uncertainty in neutron response is evaluated by changing the neutron inelastic 
cross section within experimental uncertainties~\cite{Abfalterer:2001gw,Schimmerling:1973bb,
Voss:1956,Slypen:1995fm,Franz:1989cf,Tippawan:2008xk,Bevilacqua:2013rfq,Zanelli:1981zz}
through event reweighting. The reweighting is applied to the leading neutron in the event. 
A large fraction of the secondary $\pi^0$ in the background is estimated to arise from $\pi^{-} \rightarrow \pi^0$
CEX, for which the cross sections are poorly known. The effect of this uncertainty on our 
measurement is evaluated by changing the CEX cross section within its 
uncertainty of $\pm$50\%~\cite{Ashery:1981tq,Bowles:1981,Jones:1993}, and then 
re-measuring the cross sections. Finally, the electromagnetic energy scale uncertainty (2.2\%) 
is taken from the fitted mean uncertainty of the data $m_{\gamma\gamma}$ distribution. 
Flux-integrated single differential cross sections in $\pi^0$ momentum and angle with 
statistical, systematic, and total uncertainties are summarized in 
Tables.\footnoteremember{footsupp1}{See Supplemental Material\ \SuppLocation}

\section{Results and discussion}

\begin{figure}[t]
\centering
\ifnum\PLsupp=0
  \includegraphics[width=1.0\columnwidth]{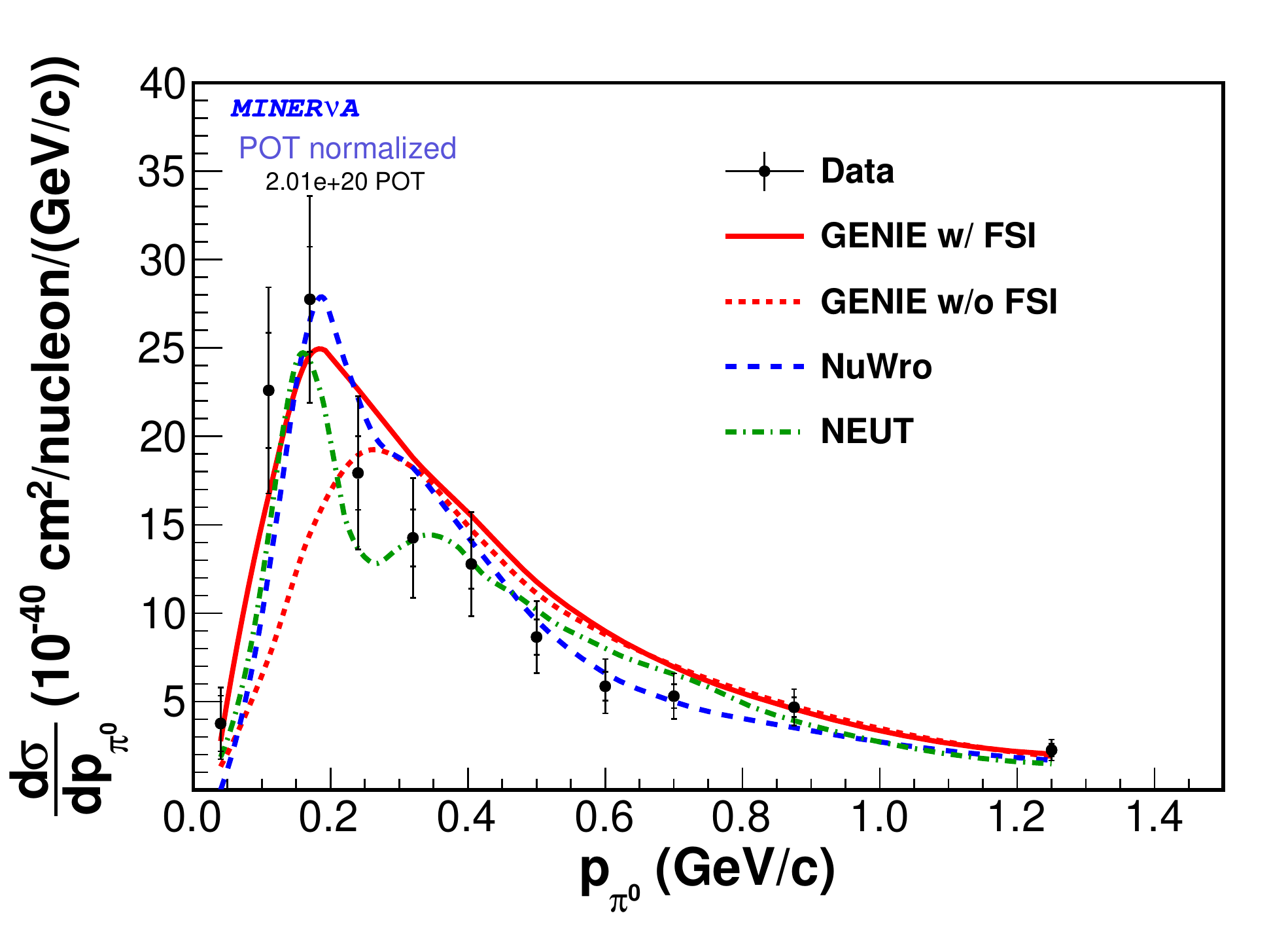}
\else
  \includegraphics[width=1.0\columnwidth]{pimom-data-mc-xs-multiflux-model-nuwro-pot.pdf}
\fi
\vspace{-7pt}
\caption{Differential cross section for $1\pi^0$ production as function of $\pi^0$ momentum. 
Data are shown as solid circles. 
The inner (outer) error bars correspond to statistical (total) uncertainties. 
The solid (dashed) histograms are 
GENIE prediction with (without) FSI, the long-dashed histogram is the prediction from the NuWro generator, 
and the dot-dashed histogram is the prediction from NEUT.}

\label{fig:xsec-pimom}
\end{figure}

\begin{figure}[!ht]
\centering
\ifnum\PLsupp=0
  \includegraphics[width=1.0\columnwidth]{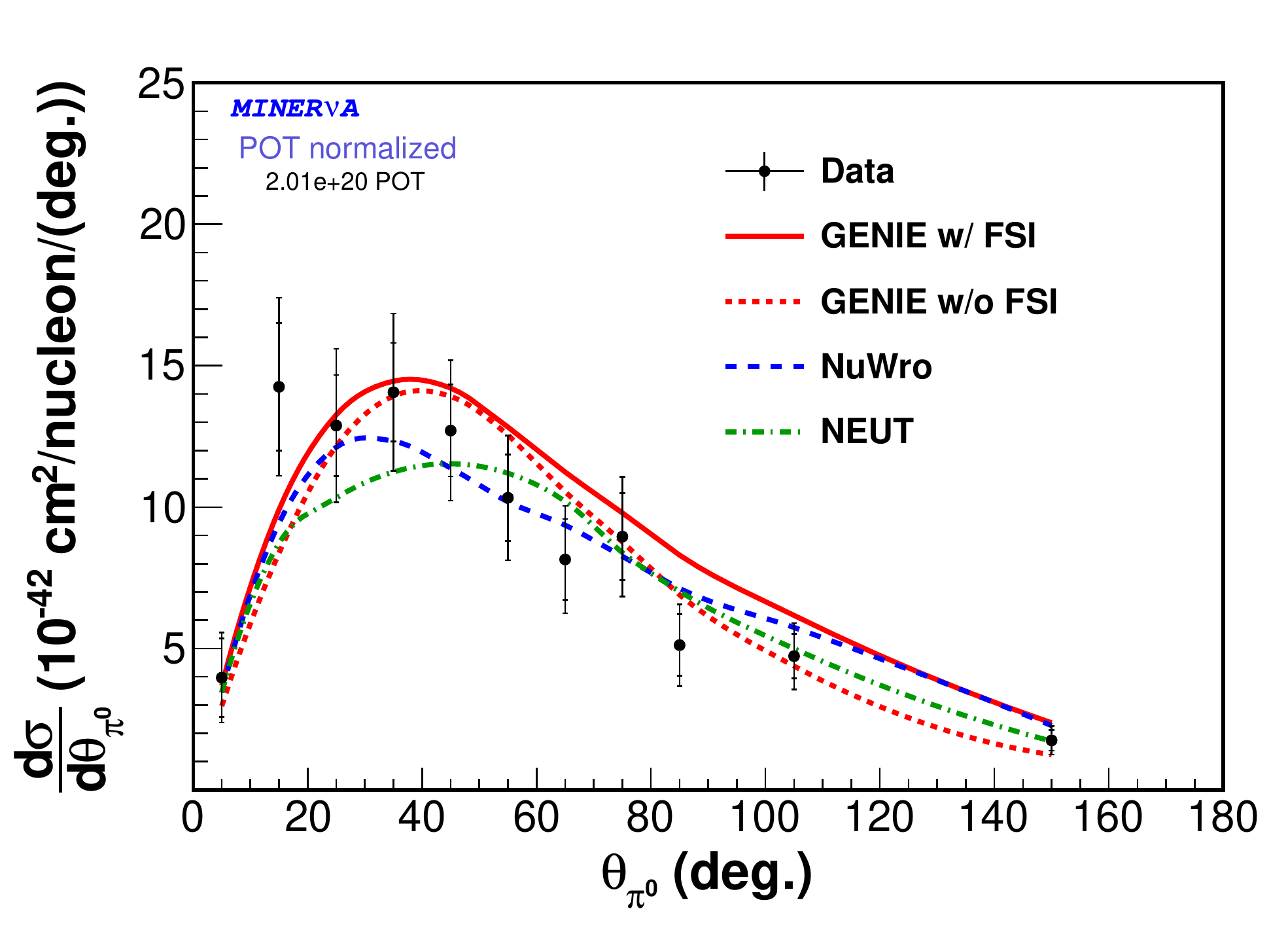}
\else
  \includegraphics[width=1.0\columnwidth]{theta-data-mc-xs-multiflux-model-nuwro-pot.pdf}
\fi
    \vspace{-7pt}
\caption{Differential cross section for $1\pi^0$ production as function of the $\pi^0$ polar angle. 
Data are shown as solid circles. The inner (outer) error bars correspond to statistical (total) 
uncertainties. The solid (dashed) histograms are GENIE prediction with (without) FSI, the 
long-dashed histogram is the prediction from the NuWro generator,
and the dot-dashed histogram is the prediction from NEUT.}
\label{fig:xsec-pitheta}
\end{figure}

The measured differential cross sections as function of $\pi^0$ 
momentum and angle with respect to the beam direction are shown in 
Figures.~\ref{fig:xsec-pimom} and \ref{fig:xsec-pitheta}, respectively. 
The data are compared to the predictions from GENIE with and without FSI.
Above $\unit[0.7]{GeV/c}$, FSI effects have little influence on the 
$\pi^0$ momentum distribution, and both predictions are in good agreement 
with the data.  For momenta below $\unit[0.3]{GeV/c}$, 
inclusion of FSI gives an increased and shifted cross section relative to the no FSI case.
This trend is exhibited by the data; GENIE calculations with and without 
FSI give a $\chi^2$ of $25.4$ and $50.0$ for 11 degrees of freedom (dof), respectively.  
For the distribution of $\pi^0$ production angle in Fig.~\ref{fig:xsec-pitheta}, 
inclusion of FSI into the GENIE simulation results in a mild flattening of the 
distribution with no significant improvement in the $\chi^2$ compared to the 
no FSI case, 16.7 versus 16.0 for 11 degrees of freedom.  Thus the effects of FSI are more pronounced with respect to 
$\pi^0$ momenta than with $\pi^0$ production angle.  This situation likely reflects 
the influence of the $\Delta(1232)$ resonance, which gives a particularly strong 
momentum dependence to the pion-nucleus interaction for 
$p_\pi \approx 0.26$ GeV/c where the pion-carbon cross section is maximum.

Figures.~\ref{fig:xsec-pimom} and \ref{fig:xsec-pitheta} also show 
predictions including FSI from the NuWro and NEUT event generators.
Any prediction requires knowledge of $\bar{\nu}_\mu N \rightarrow \mu^+ \pi^0 N$ reactions.  
Because of the dearth of data for these channels, the calculations use isospin relations 
to extrapolate from other pion production channels~\cite{Rein:1980wg,Graczyk:2009qm}. 
Both NuWro and NEUT use full cascade models~\cite{Salcedo:1987md} for the FSI description. 
The $\chi^2/$dof values for the NuWro and NEUT comparisons are 25.0/11 
and 24.9/11  for the $\pi^0$ momentum and 11.8/11 and 14.0/11 for the 
$\pi^0$ production angle, respectively. These $\chi^2$ values indicate that a 
common level of agreement is achieved by all three generators 
for $\pi^0$ momentum, and that modest differences with predictions 
versus data are observed for $\pi^0$ production angle.

There are uncertainties in the FSI used in all calculations.  Each assumes the pion FSI
after production in the nuclear medium is the same as for pion beams; this assumes
no medium effects beyond Fermi momentum and a simple binding energy.  Since
$\pi^0$ beams are not possible, isospin relations must be used for $\pi^0$ FSI.
Experiments like these are valuable for testing these approximations.  In spite of
these uncertainties, the calculations give adequate descriptions of these new data.

\begin{figure}[t]
\centering
\ifnum\PLsupp=0
  \includegraphics[width=1.0\columnwidth]{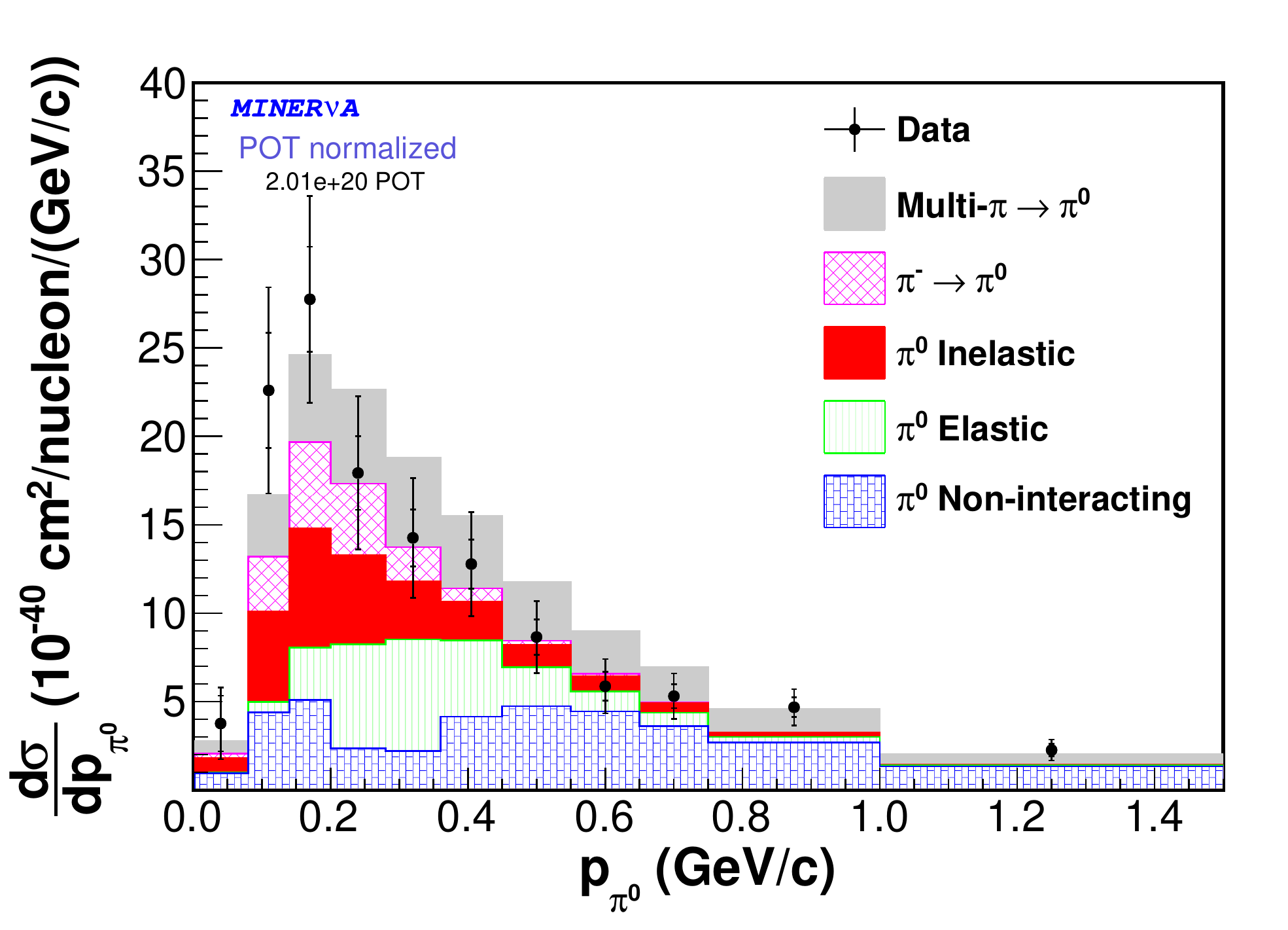}
\else
  \includegraphics[width=1.0\columnwidth]{pimom-data-mc-xs-multiflux-origin-breakdown-pot.pdf}
\fi
    \vspace{-7pt}
\caption{Same cross section data as Fig. \ref{fig:xsec-pimom}.
The stacked histograms 
show a decomposition of the 1$\pi^0$ signal into different FSI channels as predicted by GENIE. 
Description of the FSI channels follows (from top to bottom):
1) Multi-$\pi\rightarrow\pi^0$: multi-pion produced by the primary interaction and all other pions are re-absorbed
inside the nucleus, except a $\pi^0$,
2) $\pi^-\rightarrow\pi^0$: a $\pi^-$ produced by the primary interaction, then charge exchanges inside the nucleus, 
3) $\pi^0$ produced by the primary interaction and then undergoing inelastic scattering inside the nucleus,
4) $\pi^0$ produced by the primary interaction and then undergoing elastic scattering inside the nucleus, and
5) $\pi^0$ produced by the primary interaction and exiting the nucleus without interacting.}
\label{fig:breakdown-xsec-pimom}
\end{figure}

\begin{figure}[t]
\centering
\ifnum\PLsupp=0
  \includegraphics[width=1.0\columnwidth]{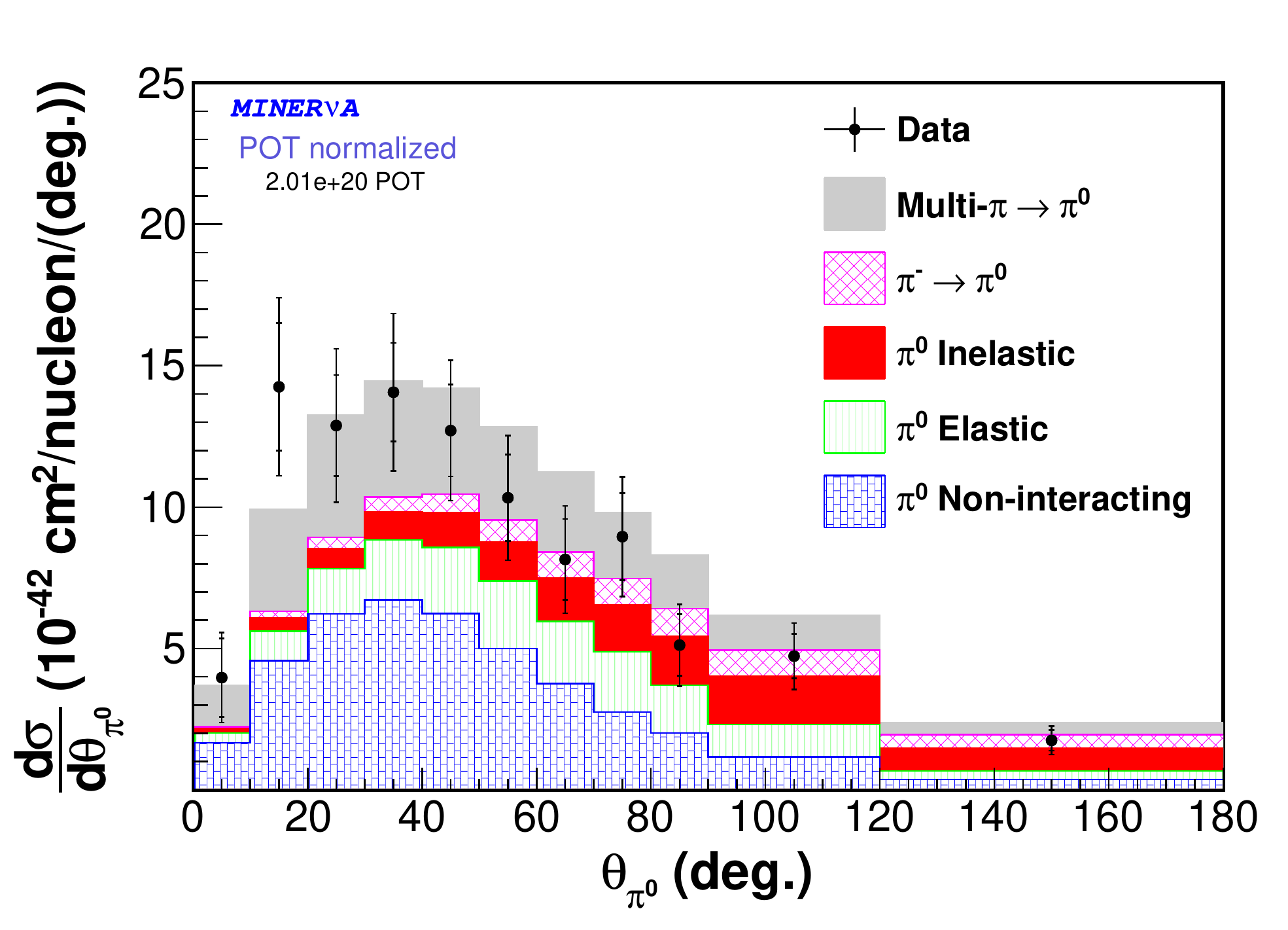}
\else
  \includegraphics[width=1.0\columnwidth]{theta-data-mc-xs-multiflux-origin-breakdown-pot.pdf}
\fi
    \vspace{-7pt}
\caption{Same cross section data as Fig. \ref{fig:xsec-pitheta} and same decomposition as with Fig.~\ref{fig:breakdown-xsec-pimom}. 
The stacked histograms show a decomposition of the 1$\pi^0$ signal into different FSI channels as predicted by GENIE. 
}
\label{fig:breakdown-xsec-pitheta}
\end{figure}

Figures.~\ref{fig:breakdown-xsec-pimom} and \ref{fig:breakdown-xsec-pitheta} 
show decompositions of the $\pi^0$ momentum and angle spectra of
Figures.~\ref{fig:xsec-pimom} and \ref{fig:xsec-pitheta} according 
to the FSI channels as predicted by the GENIE simulation, allowing for a more detailed 
interpretation.  This shows how each FSI process changes the spectrum and gives
an indication of what refinements are needed for better agreement with data.
GENIE predicts that about 80\% of the events undergo some FSI.  In the momentum spectrum, absorption events (not shown 
in the spectrum because the pion disappears in the nucleus) 
preferentially deplete the region centered at $p_\pi \approx 0.26$ GeV/c, the momentum where the
pion-carbon total reaction cross section is a maximum~\cite{Lee:2002eq}.  On the other
hand, pion inelastic 
scattering and CEX interactions shift the pion momentum from high values to 
lower values, therefore contributing to the buildup of events at 
$p_\pi \approx 0.2$ GeV/c seen in the data.  The contributions from multi-pion production followed 
by absorption, elastic scattering, and the non-interacting
component have no noticeable effect on shape.  

For the angle spectrum FSI decomposition (Fig.~\ref{fig:breakdown-xsec-pitheta})
the effects are very different.  The interactions displayed in the figure
have a small influence on shape and the angular distribution of
the $\Delta(1232)$ decay becomes important.  The inelastic and
CEX interactions are largely responsible for the buildup of events
at backward angles.

Comparison of the observed spectral shape in Fig.~\ref{fig:breakdown-xsec-pimom} to the GENIE prediction suggests
that an increase in inelastic scattering together with compensating reduction in elastic 
and/or multiple-pion production with absorption, could improve the agreement with data.
On the other hand,  Fig.~\ref{fig:breakdown-xsec-pitheta} indicates such changes would worsen the agreement with 
the data  for pion production at backwards angles.  Thus it appears that a refined 
description might require separate, independent adjustments to the two spectra.

While the previously reported information on the reaction studied here is too limited to 
enable comparisons, it is useful to compare with the recently reported MINERvA observations 
on $\nu_\mu$ induced charged pion production~\cite{Eberly:2014mra}. The latter data was obtained using a 
low-energy exposure in the same beamline, and the cross section normalizations are carried out 
in a similar way for both analyses.  Existence of both results provides stronger
constraints on the calculations. 
The $\pi^0$ momentum range in this analysis
is wider than the range shown for the charged pions of Ref.~\cite{Eberly:2014mra} (0-1.5 GeV/c versus 0.1-0.5 GeV/c) 
because the $\pi^+$ data is limited by tracking and particle identification requirements. 
Nevertheless, both analyses show that GENIE predictions are significantly improved when FSI are accounted for.   
Both results show a peak in the momentum distribution at $p_\pi \approx 0.2$ GeV/c which is seen in the GENIE calculations, 
but do not have the correct distribution at energies near this peak.
The GENIE predictions for the absolute rates for singly-produced pions exceed the observed 
rates in both analyses, however the discrepancy is less severe for production by
$\bar{\nu}_\mu$ as reported here.

\section{Conclusion}

The single differential cross sections in $\pi^0$ momentum 
and angle have been measured for 1$\pi^0$ production by $\bar{\nu}_\mu$ charged-current 
interactions in plastic scintillator (CH). 
The measurements are found to be in agreement with the predictions from 
GENIE, NuWro, and NEUT event generators. This agreement is interesting because of 
the approximations needed to make predictions
for this channel.  The measured cross section in $\pi^0$ momentum disfavors 
the GENIE prediction without FSI effects at low $\pi^0$ momentum, which is not surprising 
since FSI is expected for hadrons inside nuclei. A decomposition 
of the FSI effects shows that inelastic scattering and CEX reactions
are responsible for the peak at $p_\pi\approx 0.2$ GeV/c. These contributions
could be adjusted within external experimental errors to achieve
even better agreement of calculations with the $\pi^0$ momentum data.
However, these changes would not be as effective for the $\pi^0$ polar angle. 

Charged-current single pion production in the few GeV region of neutrino energy is an important 
class of interactions in long-baseline neutrino oscillation experiments.   This work presents 
the first detailed measurements of the kinematic distributions for single $\pi^0$s produced 
in charged current $\bar{\nu}_\mu$ interactions on carbon.  These distributions provide 
constraints on antineutrino-induced $\pi^0$ backgrounds as will occur in $\bar{\nu}_e$ appearance 
experiments.

\section*{Acknowledgments}
This work was supported by the Fermi National Accelerator Laboratory
under US Department of Energy contract
No. DE-AC02-07CH11359 which included the \minerva construction project.
Construction support also
was granted by the United States National Science Foundation under
Award PHY-0619727 and by the University of Rochester. Support for
participating scientists was provided by NSF and DOE (USA) by CAPES
and CNPq (Brazil), by CoNaCyT (Mexico), by CONICYT (Chile), by
CONCYTEC, DGI-PUCP and IDI/IGI-UNI (Peru), by Latin American Center for
Physics (CLAF) and by RAS and the Russian Ministry of Education and Science (Russia).  We
thank the MINOS Collaboration for use of its
near detector data. Finally, we thank the staff of
Fermilab for support of the beamline and the detector.

\bibliographystyle{elsarticle-num}
\section*{References}
\bibliography{ccpizero}

\ifnum\PLsupp=0
  \clearpage
    
\newcommand{\qsq}{\ensuremath{Q^2_{QE}}\xspace}
\renewcommand{\textfraction}{0.05}
\renewcommand{\topfraction}{0.95}
\renewcommand{\bottomfraction}{0.95}
\renewcommand{\floatpagefraction}{0.95}
\renewcommand{\dblfloatpagefraction}{0.95}
\renewcommand{\dbltopfraction}{0.95}
\setcounter{totalnumber}{5}
\setcounter{bottomnumber}{3}
\setcounter{topnumber}{3}
\setcounter{dbltopnumber}{3}

\ifdefined\PLsupp
  \ifnum\PLsupp=0
    \newcommand{\SuppName}{Appendix}
  \else
    \newcommand{\SuppName}{supplemental material}
  \fi
\else
  \newcommand{\SuppName}{supplemental material}
\fi

\section{\SuppName}
This\ \SuppName\ contains additional tables that are referenced in the article.

\begingroup
\begin{table}[h]
\centering
\caption{ Flux-averaged differential cross section in $\pi^0$ momentum, $d\sigma/dp_{\pi^0}(10^{-40}\text{cm}^2/\text{nucleon}/(\text{GeV/c})$, for 1$\pi^0$ production with statistical (stat), systematic (sys), and total (tot) uncertainties. }
\label{tab:xs-pimom}
\resizebox{\linewidth}{!}{
\begin{tabular}{crccccc}

\hline\hline
~$p_{\pi^0}$ (GeV/c) &~ $d\sigma/dp_{\pi^0}$ ~&~ stat(\%) &~ sys(\%) &~ tot(\%)  \\
\hline
0.00 - 0.08  &  3.75    &  41    &  33  &  53\\ 
0.08 - 0.14  & 22.60    &  14    &  21  &  25\\ 
0.14 - 0.20  & 27.75    &  10    &  18  &  21\\ 
0.20 - 0.28  & 17.92    &  11    &  21  &  24\\ 
0.28 - 0.36  & 14.26    &  11    &  20  &  23\\ 
0.36 - 0.45  & 12.77    &  10    &  20  &  22\\ 
0.45 - 0.55  &  8.65    &  11    &  20  &  23\\ 
0.55 - 0.65  &  5.86    &  13    &  22  &  26\\ 
0.65 - 0.75  &  4.67    &  12    &  20  &  24\\ 
0.75 - 1.00  &  2.25    &  11    &  18  &  21\\ 
1.00 - 1.50  &  1.27    &  15    &  21  &  26\\ 
\hline
\hline
\end{tabular}
}
\end{table}
\endgroup

\begingroup
\begin{table}[h]
\centering
\caption{ Flux-averaged differential cross section in $\pi^0$ angle, 
$d\sigma/d\theta_{\pi^0}(10^{-42}\text{cm}^2/\text{nucleon}/\text{deg.})$, for 1$\pi^0$ production 
with statistical (stat), systematic (sys), and total (tot) uncertainties. }
\label{tab:xs-theta}
\resizebox{\linewidth}{!}{
\begin{tabular}{crccccc}

\hline\hline
~$\theta_{\pi^0}$ (deg.) &~ $d\sigma/d\theta_{\pi^0}$~ &~ stat(\%) &~ sys(\%) &~ tot(\%)  \\
\hline
0  - 10  &  3.97   &   34    &  19  & 40 \\ 
10 - 20  & 14.25   &   15    &  15  & 22 \\ 
20 - 30  & 12.88   &   13    &  15  & 21 \\ 
30 - 40  & 14.06   &   12    &  15  & 19 \\ 
40 - 50  & 12.70   &   12    &  15  & 19 \\ 
50 - 60  & 10.32   &   14    &  15  & 21 \\ 
60 - 70  &  8.14   &   17    &  15  & 23 \\ 
70 - 80  &  8.95   &   17    &  16  & 23 \\ 
80 - 90  &  5.11   &   21    &  18  & 28 \\ 
90 - 120 &  4.73   &   16    &  18  & 24 \\ 
120- 180 &  1.75   &   20    &  20  & 28 \\ 
\hline
\hline
\end{tabular}
}
\end{table}
\endgroup

\clearpage

\fi

\end{document}